\documentclass[conference,letterpaper]{IEEEtran}

\usepackage{graphicx,psfrag,epsfig,epsf,latexsym,hhline,amsmath,amssymb,multirow}
\usepackage{pst-plot}
\usepackage{pstricks-add}
\usepackage{pifont}
\usepackage{graphicx}
\usepackage{amsthm}
\usepackage[linesnumbered,ruled,vlined]{algorithm2e}
\usepackage[noadjust]{cite}
\usepackage{array}
\usepackage{blindtext}
\usepackage{etoolbox}
\usepackage{comment}
\usepackage{epstopdf}
\usepackage{hyperref}
\usepackage{slashbox}
\usepackage{bbm}
\usepackage{arydshln}
\usepackage{lipsum}

\newcommand\xqed[1]{%
  \leavevmode\unskip\penalty9999 \hbox{}\nobreak\hfill
  \quad\hbox{#1}}
\newcommand\demo{\xqed{$\blacksquare$}}

\SetCommentSty{mycommfont}

\SetKwInput{KwInput}{Input}                
\SetKwInput{KwOutput}{Output}              

\newcolumntype{M}[1]{>{\centering\arraybackslash}m{#1}}
\newcolumntype{P}[1]{>{\centering\arraybackslash}p{#1}}

\newcommand{\iid}{i.\@i.\@d.\ }

\theoremstyle{definition}

\newtheorem{example}{Example}
\newtheorem{theorem}{Theorem}
\newtheorem{lemma}{Lemma}

\IEEEoverridecommandlockouts
\begin{document}
\title{Sub-Block Rearranged Staircase Codes for Optical Transport Networks}
\author{
Min Qiu and Jinhong Yuan\\
University of New South Wales, Sydney, Australia\\
\{min.qiu, j.yuan\}@unsw.edu.au

\thanks{This work was supported in part by the Australian Research Council (ARC) Discovery Project under Grant DP220103596, and in part by the ARC Linkage Project under Grant LP200301482.
}%
}

\maketitle

\begin{abstract}
We propose a new family of spatially coupled product codes, called sub-block rearranged staircase (SR-staircase) codes. Each SR-staircase code block is constructed by encoding rearranged preceding code blocks and new information blocks, where the rearrangement involves sub-blocks decomposition and transposition. The proposed codes can be constructed to have each code block size of $1/q$ to that of the conventional staircase codes while having the same rate and component codes, for any positive integer $q$. In this regard, we can use strong algebraic component codes to construct SR-staircase codes with a similar or the same code block size and rate as staircase codes with weak component codes. Moreover, both waterfall and error floor performance can be further improved by using a large coupling width. The superior performance of the proposed codes is demonstrated through density evolution and error floor analysis as well as simulation.
\end{abstract}

\section{Introduction}\label{sec:int}
Modern optical transport networks (OTN) are required to support data transmission of 400 Gbit/s and beyond over long distances. As a result, the use of most packet retransmission protocols become inefficient. This necessitates the design of low-complexity forward error correction (FEC) coding schemes that can achieve a rate close to capacity while having extremely low error floor. Particularly, a bit error rate (BER) lower than $10^{-15}$ is required in the error floor region \cite{6805157,optical_FEC_2020}. Spatially coupled low-density parity-check codes \cite{782171} have become the popular candidates for OTN \cite{7340116} due to their close-to-capacity performance and low error floor \cite{5571910,5695130}. However, their remarkable performance requires soft-decision decoding which poses several challenges in implementation, such as large internal data flow \cite{Smith12} and high hardware and power cost for enabling high-resolution analog-to-digital conversion.

The FEC codes based on hard-decision decoding have significantly lower decoding complexity and hardware costs, which makes them appealing to OTN. The authors in \cite{Smith12} proposed staircase codes by applying spatial coupling to product codes \cite{6917198} with Bose-Chaudhuri-Hocquengham (BCH) component codes. Staircase codes can achieve a performance within 0.56 dB from the capacity of the binary symmetric channel (BSC) under iterative bounded distance decoding (iBDD) \cite{optical_FEC_2020} and outperform existing FEC solutions in ITU-T G.975.1 \cite{G7951}. Another class of spatially coupled product codes called braided block codes were introduced in \cite{4957627}, which, when using BCH component codes, i.e., braided BCH codes \cite{6831429}, have comparable error performance to staircase codes. Both codes \cite{Smith12,6831429} can be described under a unified framework named ``zipper codes'' \cite{8929906}. The authors in \cite{8929906} also proposed tiled diagonal zipper codes which can be seen as a combination of continuously interleaved BCH codes \cite{Coe2012} and staircase codes \cite{Smith12}.

In addition to spatial coupling, another line of work is to construct symmetry-based product codes \cite{7309002} to reduce the blocklength of product codes \cite{1057464} while having the same component code and similar code rates. Thanks to this property, one can also employ stronger algebraic component codes to construct symmetry-based product codes in a bid to achieve better waterfall and error floor performance while maintaining similar blocklengths and code rates as the conventional product codes. The first examples of such codes are half-product codes \cite{5671556}, whose codewords are derived from product codes with the additional constraint that the code arrays are anti-symmetric, thereby leading to an effective blocklength about half to that of the product codes. Later, this idea motivated the design of quarter-product codes and octal-product codes in \cite{7282455} as well as half-braided BCH codes in \cite{7537380}. However, all the above symmetric-based product codes require square code blocks and the same component codes for row and column encoding. In addition, the codes in \cite{7282455} restrict the component codes to be reversible (i.e., a code that is invariant under a reversal of the coordinates in each codeword \cite{reversible1}). These restrictions reduce the design space of symmetric-based product codes and may limit their potential applications.

This paper focuses on designing new FEC schemes under low-complexity iBDD to achieve better waterfall and error floor performance with lower miscorrection probability than staircase codes \cite{Smith12} with similar blocklengths and rates. Motivated by spatial coupling and symmetry, we propose sub-block rearranged staircase (SR-staircase) codes. The proposed codes can be constructed to have each code block with a size of $1/q$ to that of the conventional staircase codes with the same algebraic component codes while maintaining the same code rate, for any positive integer $q$. This means that we can employer strong algebraic component codes to construct SR-staircase codes with a similar or the same code block size and rate as staircase codes with weak component codes. However, unlike all the aforementioned symmetric-based product codes, the proposed codes do not impose any additional constraint on the component codes and code array shapes. We use density evolution to characterize the decoding thresholds on the BSC and investigate the error floor by analyzing the contributing error patterns. Both theoretical and simulation results show that the designed SR-staircase codes achieve better waterfall and error floor performance over the conventional staircase codes under iBDD. Moreover, the decoding threshold and error floor of SR-staircase codes can be further improved by using a large coupling width.

This paper uses the following notations. Let $\mathbb{N}$ represent the set of natural numbers. The sets of even and odd natural numbers are represented by $2\mathbb{N}$ and $2\mathbb{N}-1$, respectively. We define $[n] \triangleq \{1,\ldots,n\}$ for any $n\in \mathbb{N}$. $\left\lceil x\right\rceil$ gives the nearest integer that is not less than $x$. The indicator function is represented by $\mathbbm{1}\{\cdot\}$.


%


\section{Sub-block Rearranged Staircase Codes}

We consider that the underlying component codes are binary primitive BCH codes. A SR-staircase code comprises a sequence of code blocks $\boldsymbol{B}_1,\boldsymbol{B}_2\ldots.$ At time $i\in \mathbb{N}$, code block $\boldsymbol{B}_i = [\boldsymbol{K}_i,\boldsymbol{P}_i]$ is a concatenation of information block $\boldsymbol{K}_i$ and parity block $\boldsymbol{P}_i$. To construct the SR-staircase code, two shortened BCH codes $\mathcal{C}_j$ for $j\in\{1,2\}$ are used. We denote by $k_j$, $n_j$, $t_j$, $e_j$, and $\boldsymbol{G}_j$ the message length, codeword length, error correction capability, shortening parameter, and generator matrix, respectively, of $\mathcal{C}_j$. Note that we can also express the codeword length and information length of $\mathcal{C}_j$ as $n_j = 2^{\nu_j}-1-e_j$ and $k_j = 2^{\nu_j}-1-\nu_j t_j-e_j$, respectively, for some positive integers $\nu_j \geq 3$, where $\nu_j$ is Galois field extension \cite[Chap. 3.3]{Lin09}.

\subsection{Construction}
The encoding of SR-staircase codes is performed in a recursive manner like the conventional staircase codes. The main difference is that each preceding SR-staircase code block $\boldsymbol{B}_{i-1}$ is decomposed into $q_1$ equal-size sub-blocks if $i\in 2\mathbb{N}-1$ and $q_2$ equal-size sub-blocks if $i\in 2\mathbb{N}$. Each sub-block is then transposed and encoded row-by-row with BCH encoding to obtain the current code block $\boldsymbol{B}_{i}$. The size of $\boldsymbol{B}_{i}$ is $\frac{m_1}{q_1}\times m_2$ if $i\in 2\mathbb{N}-1$ and $\frac{m_2}{q_2}\times m_1$ if $i\in 2\mathbb{N}$. Moreover, all the bits in each row of $\boldsymbol{B}_{i}$ are the last $m_2$ bits of a codeword of $\mathcal{C}_2$ when $i\in 2\mathbb{N}-1$ and the last $m_1$ bits of $\mathcal{C}_1$ when $i\in 2\mathbb{N}$. Note that the numbers of columns for $\boldsymbol{B}_i$, $m_1$ and $m_2$, have to be divisible by $q_1$ and $q_2$, respectively. We also denote by $w$ the coupling width, where $w \geq 2$ and both $m_1$ and $m_2$ have to be divisible by $w-1$. 

\subsubsection{Case $w=2$}\label{subsec:enc}
For ease of presentation, we first describe the encoding steps for $i\in 2\mathbb{N}$.

\emph{Step 1 (Initialization):} Set all the entries of $\boldsymbol{B}_0$ to zero: $\boldsymbol{B}_0 = \boldsymbol{0}_{\frac{m_2}{q_2},m_1}$, which are known by the encoder and decoder.

\emph{Step 2 (Decomposition):} The preceding block $\boldsymbol{B}_{i-1}$ with size $\frac{m_1}{q_1}\times m_2$ is divided into $q_2$ consecutive equal-size sub-blocks with size $\frac{m_1}{q_1}\times \frac{m_2}{q_2}$ as
\begin{align}\label{eq:interleaver1}
\boldsymbol{B}_{i-1} = \left[\boldsymbol{B}_{i-1,1},\boldsymbol{B}_{i-1,2},\ldots,\boldsymbol{B}_{i-1,q_2}\right].
\end{align}

\emph{Step 3 (Transposition):} Apply matrix transpose to each sub-block decomposed from $\boldsymbol{B}_{i-1}$ in Step 2 and recombine all the transposed sub-blocks into a size $\frac{m_2}{q_2} \times \frac{m_1 q_2}{q_1}$ block $\boldsymbol{B}^{\pi}_{i-1}$
\begin{align}\label{eq:int1}
\boldsymbol{B}^{\pi}_{i-1}=\left[\boldsymbol{B}^\mathsf{T}_{i-1,1},\boldsymbol{B}^\mathsf{T}_{i-1,2},\ldots,\boldsymbol{B}^\mathsf{T}_{i-1,q_2}\right].
\end{align}
Each sub-block $\boldsymbol{B}^{\pi}_{i-1,l}$ has a size of $\frac{m_2}{q_2}\times\frac{m_1}{q_1}$ for $l\in [q_2]$. Note that all bits in the same column position of every transposed sub-block, $\boldsymbol{B}^{\pi}_{i-1,1},\ldots,\boldsymbol{B}^{\pi}_{i-1,q_2}$, belong to the same BCH component codeword of $\mathcal{C}_2$.

\emph{Step 4 (Array Concatenation):} Construct the message matrix with size $\frac{m_2}{q_2}\times (k_1+e_1)$ to be encoded at time $i$,
\begin{align}\label{eq:K_prime1}
\boldsymbol{K}'_i = \left[\boldsymbol{0}_{\frac{m_2}{q_2},e_1},\boldsymbol{B}^{\pi}_{i-1},\boldsymbol{K}_i\right],
\end{align}
where $\boldsymbol{K}_i$ is an $\frac{m_2}{q_2}\times (k_1-\frac{m_1q_2}{q_1})$ block filled with information bits, and $\boldsymbol{0}_{\frac{m_2}{q_2},e_1}$ represents the block filled with shortened bits.

\emph{Step 5 (Component Code Encoding):} Obtain the codeword matrix with size $\frac{m_2}{q_2}\times n_1$ at time $i$ as
\begin{align}
\boldsymbol{C}_i =& \boldsymbol{K}'_i\boldsymbol{G}_1  \nonumber \\ =&\left[\boldsymbol{0}_{\frac{m_2}{q_2},e_1},\boldsymbol{B}^{\pi}_{i-1},\boldsymbol{K}_i,\boldsymbol{P}_i\right]\nonumber \\
=&\left[\boldsymbol{0}_{\frac{m_2}{q_2},e_1},\boldsymbol{B}^{\pi}_{i-1},\boldsymbol{B}_i\right],
\end{align}
where $\boldsymbol{P}_i$ is the parity block with size $\frac{m_2}{q_2}\times (n_1-k_1)$, and $\boldsymbol{B}_{i}=$
$[\boldsymbol{K}_i,\boldsymbol{P}_i]$ is an $\frac{m_2}{q_2}\times m_1$ code block to be transmitted. Each row of $[\boldsymbol{B}^{\pi}_{i-1},\boldsymbol{B}_i]$ is a shortened codeword of $\mathcal{C}_1$.

The steps to obtain $\boldsymbol{B}_i$ for $i\in 2\mathbb{N}-1$ are similar to the above. After Step 5, each row of $[\boldsymbol{B}^{\pi}_{i-1},\boldsymbol{B}_i]$ is a shortened codeword of $\mathcal{C}_2$ for $i\in 2\mathbb{N}-1$. For $\boldsymbol{B}_i$, the relation between the component codeword length, time index $i$, the number of decomposed sub-blocks, and the number of columns satisfies
\begin{align}\label{eq:con1}
n_{\varphi(i)} = m_{\varphi(i)}+\frac{m_{\varphi(i)}\cdot q_{\varphi(i-1)}}{q_{\varphi(i)}},
\end{align}
where $\varphi(.)$ is a mapping function such that $\varphi(x) = \frac{3-(-1)^x}{2}$.

Finally, Steps 1-5 are performed for $i\geq 1$ to obtain all code blocks. The code rate is
\begin{align}
R =& \frac{1}{2}\left(\frac{k_1}{m_1}+\frac{k_2}{m_2}-\frac{q_2}{q_1}-\frac{q_1}{q_2}\right) \nonumber \\
=&1-\frac{1}{2}\left(\frac{\nu_1t_1}{m_1}+\frac{\nu_2t_2}{m_2}\right). \nonumber 
\end{align}

Note that in Step 3, the transformation of $\boldsymbol{B}_{i-1}$ into $\boldsymbol{B}^{\pi}_{i-1}$ in \eqref{eq:int1} can be generalized by adding a permutation function $\pi(.)$ which permutes the rows and columns of a matrix, i.e.,
\begin{align}\label{enc_random_int}
\boldsymbol{B}^{\pi}_{i-1}=\pi\left(\left[\boldsymbol{B}^\mathsf{T}_{i-1,1},\boldsymbol{B}^\mathsf{T}_{i-1,2},\ldots,\boldsymbol{B}^\mathsf{T}_{i-1,q_2}\right]\right).
\end{align}
Alternatively, Step 3 may be described by using the zipper code framework \cite{8929906} by specifying a bijective interleaver map which maps the position of each bit from $\boldsymbol{B}_{i-1}$ to $\boldsymbol{B}^{\pi}_{i-1}$.

\subsubsection{Case $w>2$}\label{subsec:enc_w3}
In this case, we need to ensure that each sub-block used for coupling has the same size. This is possible if and only if $m_1=m_2 \triangleq m$ and $q_1 = q_2 \triangleq q$. Consider $i\in2\mathbb{N}$. To obtain $\boldsymbol{B}_i$, we first modify Step 1 of the encoding in Sec. \ref{subsec:enc} by setting $\boldsymbol{B}_0,\ldots,\boldsymbol{B}_{w-2}$ to all-zero matrices. Next, we modify Step 4 by further dividing the transformed preceding code block $\boldsymbol{B}^{\pi}_{i-l}$ obtained from \eqref{eq:int1} into $w-1$ consecutive equal-size sub-blocks for $l\in[w-1]$, i.e., $\boldsymbol{B}^{\pi}_{i-l} = \left[\boldsymbol{B}^{\pi}_{i-l,1},\ldots,\boldsymbol{B}^{\pi}_{i-l,w-1}\right]$,
where each sub-block is an $\frac{m}{q}\times \frac{m}{w-1}$ binary matrix. Then, the message matrix to be encoded at time $i$ is constructed by taking the $l$-th sub-block of preceding transformed code block $\boldsymbol{B}^{\pi}_{i-l}$ for $l\in[w-1]$, i.e., $\boldsymbol{K}'_i = \left[\boldsymbol{0}_{\frac{m}{q},e_1},\boldsymbol{B}^{\pi}_{i-1,1},\boldsymbol{B}^{\pi}_{i-2,2},\ldots,\boldsymbol{B}^{\pi}_{i-w+1,w-1},\boldsymbol{K}_i\right]$.
The rest of the encoding steps are the same as those in Sec. \ref{subsec:enc}. 

It is important to note that when $w\geq q+1$, the bits in different column positions of the coupled block $[\boldsymbol{B}^{\pi}_{i-1,1},\ldots,\boldsymbol{B}^{\pi}_{i-w+1,w-1}]$ in $\boldsymbol{K}'_i$ are protected by different component codewords because any pair of sub-blocks, $\boldsymbol{B}^{\pi}_{i-l,l}$ and $\boldsymbol{B}^{\pi}_{i-l',l'}$ with $l\neq l'$ and $l,l' \in [w-1]$, are decomposed from different preceding code blocks.

\subsection{Connections to Other Spatially Coupled Codes}
SR-staircase codes are motivated and derived by introducing symmetry in the conventional staircase codes \cite{Smith12}. Consider a SR staircase code with $w=2$ and let $q \triangleq q_1 = q_2$. By concatenating $q$ identical SR-staircase code blocks $\boldsymbol{B}_i$, one obtains the resultant staircase code block as $\boldsymbol{B}^*_i = [\boldsymbol{B}^\mathsf{T}_i,\ldots,\boldsymbol{B}^\mathsf{T}_i]^\mathsf{T}$. $\boldsymbol{B}^*_i$ is obtained by encoding rearranged preceding block $\boldsymbol{B}^{\pi*}_{i-1} = [(\boldsymbol{B}^{\pi}_{i-1})^\mathsf{T},\ldots,(\boldsymbol{B}^{\pi}_{i-1})^\mathsf{T}]^\mathsf{T}$,
where the construction of $\boldsymbol{B}^{\pi}_{i-1}$ follows either \eqref{eq:int1} or \eqref{enc_random_int}. As a result, each row of $[\boldsymbol{B}^{\pi*}_{i-1},\boldsymbol{B}_{i}]$ is a valid codeword of $\mathcal{C}_1$ when $i \in 2\mathbb{N}$ and $\mathcal{C}_2$ when $i \in 2\mathbb{N}-1$. Clearly, each code block $\boldsymbol{B}^*_i$ is drawn from a subset of the set of the conventional staircase code blocks due to symmetry, i.e., having $q-1$ replicas of $\boldsymbol{B}_i$. Thus, the staircase code $\boldsymbol{B}^*_1,\ldots$ is a subcode of the conventional staircase code. Notice that when $q=1$, the encoding steps in Sec. \ref{subsec:enc} produce the conventional staircase codes. By removing any $q-1$ replicas of $\boldsymbol{B}_i$ as they do not contain any new information, the resultant SR-staircase codes achieve the same rates and an effective block size of $1/q$ to the staircase codes from which they are derived. In this regard, the proposed construction allows one to employ stronger BCH codes to construct SR-staircase codes with improved error performance while maintaining similar or the same block sizes and rates compared to the conventional staircase codes.

SR-staircase codes can also be seen as a generalization of the tiled diagonal zipper codes in \cite{8929906}. Specifically, one can obtain a tiled diagonal zipper code from the proposed construction by enforcing $\mathcal{C}_1=\mathcal{C}_2$, $w-1=q_1=q_2$, $m_1=m_2$, and using a specific block interleaver. However, we emphasize that the proposed codes are motivated and derived by applying the idea of symmetry-based product codes \cite{7309002,7282455} to staircase codes \cite{Smith12} starting from $w=2$ as illustrated above. Compared to tiled diagonal zipper codes, the proposed codes have a more flexible structure suitable for a wider range of applications and the design of code parameters will be justified via density evolution and error floor analysis.

It is also worth noting that the proposed construction is related to the class of partially coupled codes, i.e., \cite{PIC2020,GSCPCC2021ISIT,9491085}. This can be seen by noting that a fraction of information and/or parity bits in one code block are repeated and coupled to become a part of the input to the encoders of consecutive code blocks. All repeated bits are punctured before transmission. This allows us to employ stronger component codes to improve the overall performance of the coupled codes.

\subsection{Decoding}\label{sec:dec}
We restrict the decoding to be iBDD similar to \cite[Sec. IV-A]{Smith12} due to its simplicity and low complexity. The detailed decoding steps are omitted due to space limitations. We will see in Sec. \ref{sec:sim} that the performance of SR-staircase codes under iBDD is close to that under miscorrection-free iBDD \cite{8345919} due to the use of BCH component codes with larger $(t_1,t_2)$

\section{Decoding Threshold Analysis}\label{sec:pa}

\begin{figure}[t!]
	\centering
\includegraphics[width=3.5in,clip,keepaspectratio]{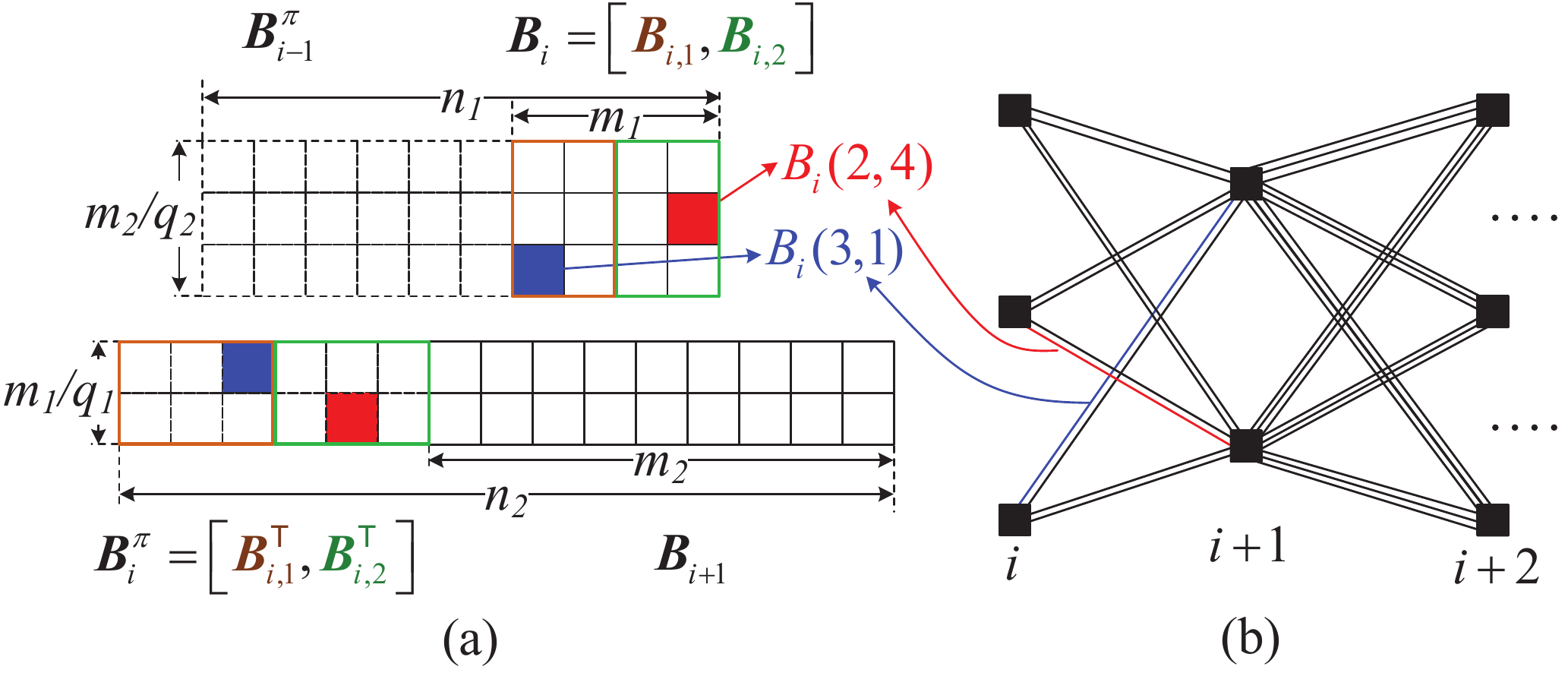}
\caption{A SR-staircase code with $w=2$, $(m_1,m_2) = (4,9)$ and $(q_1,q_2) = (2,3)$ and $i \in 2\mathbb{N}$. (a) Code blocks (dash lines and solid lines illustrate preceding and current code blocks, respectively); (b) Tanner graph.}
\label{fig:graph_w2}
\end{figure}

\subsection{Graph Model}\label{sec:graph_model}
Following the approach in \cite{7890389}, we consider a deterministic code structure since the interleaver of the proposed codes is fixed. The analysis performed on a deterministic code structure allows one to make precise statements about the performance of actual codes. 

For illustrative purpose, we consider the case of $w=2$. From Sec. \ref{subsec:enc}, we know that code block $\boldsymbol{B}_i$ has $\frac{m_2}{q_2}$ rows for $i\in 2\mathbb{N}$ and $\frac{m_1}{q_1}$ rows for $i\in 2\mathbb{N}-1$. By using the Tanner graph representation \cite{1056404}, it can be seen that the $i$-th spatial position (time instance) on the graph has $\frac{m_2}{q_2}$ check nodes (CNs) when $i\in 2\mathbb{N}$ and $\frac{m_1}{q_1}$ CNs when $i\in 2\mathbb{N}-1$ because one component codeword poses constraints on a row of $\boldsymbol{B}_i$. Each bit in $\boldsymbol{B}_i$ is represented by a variable node (VN) that connects a pair of CNs in the $i$-th and $(i+1)$-th spatial positions via an edge. Thus, each VN always has degree 2. All CNs in any two neighboring spatial positions are fully connected. More precisely, each pair of CNs in the two neighboring spatial positions $i$ and $i+1$, are connected via $q_1$ and $q_2$ edges for $i\in2\mathbb{N}$ and $i\in2\mathbb{N}-1$, respectively, where a VN lies on each edge. We use an example to illustrate the graph representation of a SR-staircase code with given specific parameters.

\begin{example}\label{exam1}
Consider a SR-staircase code with $(m_1,m_2) = (4,9)$ and $(q_1,q_2) = (2,3)$. The code blocks and the corresponding graph model of this SR-staircase code are shown in Fig. \ref{fig:graph_w2}(a) and Fig. \ref{fig:graph_w2}(b), respectively. Sub-blocks are indicated in different colors. Consider $i \in 2\mathbb{N}$. Since each VN always has degree 2, we use an edge to represent a VN that connects a pair of CNs for simplicity. We label two bits in $\boldsymbol{B}_i$, i.e., $B_i(3,1)$ and $B_i(2,4)$, in Fig. \ref{fig:graph_w2}(a) and mark their corresponding edges (VNs) in the Tanner graph with the same color in Fig. \ref{fig:graph_w2}(b). 
\demo
\end{example}

Notice that when $\min(q_1,q_2)\geq w=2$, the SR-staircase code has a multi-edge graph representation shown in Fig. \ref{fig:graph_w2}(b) such that every $q$ bits are protected by two component codewords. When $q_1=q_2=q$ and $w \geq q+1$, the SR-staircase code has a single-edge graph representation according to the construction in Sec. \ref{subsec:enc_w3}. When $2<w<q+1$, the connectivity between CNs is mixed with single-edge and multi-edge. For this case, the number of connecting edges ranges from 1 to $\lceil\frac{q}{w-1}\rceil$ and depends specifically on $(q,w)$.

\subsection{Density Evolution}
We derive the DE equations based on the graph model in \eqref{sec:graph_model}. The analysis is performed on the BSC. To make precise statements about the performance of the proposed codes with deterministic structures under iBDD, we adopt the approach in \cite{7890389} to perform DE analysis. Moreover, we assume that the iBDD is miscorrection-free according to \cite{7890389}.

We start with the case of $w=2$. Consider the SR-staircase code constructed in Sec. \ref{subsec:enc} with code blocks $\boldsymbol{B}_i,i\in[L]$. Let $p$ be the crossover probability of a BSC. We define the effect channel quality to be
\begin{align}\label{eq:MP}
M_{\varphi(i)} \triangleq p n_{\varphi(i)} \overset{\eqref{eq:con1}}{=}p \left(m_{\varphi(i)}+\frac{m_{\varphi(i)}\cdot q_{\varphi(i-1)}}{q_{\varphi(i)}}\right),
\end{align}
whose operational meaning is the expected average number of bits received in errors per component code constraint of $\mathcal{C}_{\varphi(i)}$ and $\varphi(.)$ is a mapping function defined right after \eqref{eq:con1}. We further define a parameter $x^{(\ell)}_i, i\in[L]$, whose operational meaning is that the probability of a randomly chosen erroneous bit attached to a component code of $\mathcal{C}_{\varphi(i)}$ in $\boldsymbol{B}_i$ is not recovered after $\ell$ decoding iterations converges asymptotically to $x^{(\ell)}_i$. DE is performed by tracking $x^{(\ell)}_i$. Define $f(\lambda,t)\triangleq 1-\sum_{i=1}^{t-1}\frac{\lambda^i}{i!}e^{-\lambda}$ to be the complementary Poisson cumulative distribution function for a Poisson random variable $\lambda$ with support $t$. Following \cite{7890389}, the DE equation for SR-staircase codes is
\begin{align}\label{eq:DEw2}
x^{(\ell)}_i =f\left(\frac{M_{\varphi(i)}}{2}\left(x^{(\ell)}_{i-1}+x^{(\ell-1)}_{i+1}\right),t_{\varphi(i)}\right),
\end{align}
where $x^{(0)}_i = 1$ for $i \in [L]$ and $x^{(\ell)}_i = 0$ for $i<1$ and $i>L$. Note that since $q_{\varphi(i)}$ is fixed and $n_{\varphi(i)}\gg q_{\varphi(i)}$ and based on the argument in \cite[Sec. IV-A]{7282455}, the DE equation \eqref{eq:DEw2} can be applied to the proposed codes regardless of whether the Tanner graph is single-edge or multi-edge. The BSC decoding threshold is defined as $\bar{p} \triangleq \sup\left\{p>0\left|\lim_{\ell \rightarrow \infty}\boldsymbol{x}^{(\ell)} = \boldsymbol{0}_L \right.\right\}$.

When $w > 2$, we let $m_1=m_2\triangleq m$ and $q_1=q_2 \triangleq q$ according to Sec. \ref{subsec:enc_w3}. The expected number of initial errors per component code is $M_1 = M_2 \triangleq M$. Due to coupling, the $l$-th sub-block of preceding code block $\boldsymbol{B}^{\pi}_{i-l}$ for $l\in[w-1]$ is used as a part of the inputs to encode $\boldsymbol{B}_{i}$ while $\boldsymbol{B}_{i}$ is also used as a part of the  inputs to encode $\boldsymbol{B}_{i+1},\ldots,\boldsymbol{B}_{i+w-1}$. The DE equation in \eqref{eq:DEw2} is then modified to
\begin{align}
x^{(\ell)}_i =f\left(\frac{M}{2(w-1)}\sum\nolimits_{j=1}^{w-1}\left(x^{(\ell)}_{i-j}+x^{(\ell-1)}_{i+j}\right),t_{\varphi(i)}\right).
\end{align}

\begin{table}[t!]
\setlength\tabcolsep{3pt}
  \centering
 \caption{BSC Decoding thresholds of SR-staircase Codes}\label{table2}
\begin{tabular}{c  c  c  c  c  c  c  c  c }
\hline
Scheme & Rate  & $w$ & $\nu$ & $m$ & $(t_1,t_2)$   & $q$ & $\boldsymbol{B}_i$ size & $\bar{p}$      \\\hline
\cite[Table I]{6787025} &0.941 & 2& 11 & $748$ & $(4,4)$ & $1$ & 559504 & $5.240\hspace{-0.5mm}\cdot\hspace{-1mm} 10^{-3}$   \\\hdashline
\multirow{2}{*}{Proposed} &0.941 & 2& 11 & $936$ & $(5,5)$ & $2$ & 436178 & $5.281\hspace{-0.5mm}\cdot\hspace{-1mm} 10^{-3}$   \\
 &0.941 &5&  11 & $1022$ & $(6,5)$ & $2$ & 522242 & $5.334\hspace{-0.5mm}\cdot\hspace{-1mm} 10^{-3}$   \\
  \hline
\cite[Sec. IV-C]{Smith12} &0.937 & 2& 10 & $510$ & $(3,3)$ & $1$ & 261120 & $5.630\hspace{-0.5mm}\cdot\hspace{-1mm} 10^{-3}$   \\\hdashline
\multirow{2}{*}{Proposed} &0.937 & 2& 11 & $876$ & $(5,5)$ & $3$ & 255792 & $5.643\hspace{-0.5mm}\cdot\hspace{-1mm} 10^{-3}$   \\
 &0.937 & 5& 11 & $964$ & $(6,5)$ & $4$ & 232324 & $5.655\hspace{-0.5mm}\cdot\hspace{-1mm} 10^{-3}$   \\
 \hline
\cite[Table I]{7537380}& 0.917 & 2& 10 & $360$ &$(3,3)$  & $1$ & 129600 & $7.992\hspace{-0.5mm}\cdot\hspace{-1mm} 10^{-3}$   \\\hdashline
Proposed & 0.917 & 4& 10 & $480$ &$(4,4)$  & $2$ & 115200 & $8.170\hspace{-0.5mm}\cdot\hspace{-1mm} 10^{-3}$   \\
\hline
 \cite{8345919} & 0.867 & 2& 8 & $128$ &$(2,2)$  & $1$ & 16384 & $1.402\hspace{-0.5mm}\cdot\hspace{-1mm} 10^{-2}$   \\\hdashline
Proposed & 0.867 & 4& 9 & $237$ &$(4,3)$  & $3$ & 18732 & $1.429\hspace{-0.5mm}\cdot\hspace{-1mm} 10^{-2}$   \\
 \hline
 \cite[Table I]{8856224} & 0.833 & 2& 9 & $112$ &$(2,2)$  & $1$ & 12996 & $1.574\hspace{-0.5mm}\cdot\hspace{-1mm} 10^{-2}$   \\\hdashline
\multirow{2}{*}{Proposed} & 0.833 & 5& 9 & $216$ &$(4,4)$  & $4$ & 11664 & $1.816\hspace{-0.5mm}\cdot\hspace{-1mm} 10^{-2}$   \\
& 0.834 & 5& 9 & $244$ &$(5,4)$  & $4$ & 14884 & $1.815\hspace{-0.5mm}\cdot\hspace{-1mm} 10^{-2}$   \\
 \hline
\end{tabular}
\end{table}

\subsection{Decoding Threshold Results}\label{sec:DE_result}

We use the above DE equations to compute the decoding threshold. For illustrative purpose, we consider $\nu_1=\nu_2 \triangleq \nu $, $m_1=m_2\triangleq m$, $q_1=q_2 \triangleq q$, and assume full decoding of the entire spatial code chain. The decoding thresholds of the proposed codes and some existing staircase codes are reported in Table \ref{table2}. It can be observed that the proposed codes achieve a larger threshold than the conventional staircase codes for the same or similar rates and with comparable block sizes. Moreover, increasing the coupling width provides further performance gain. Recall that all the thresholds are based on the assumption of using miscorrection-free decoding. Hence, the actual performance gain of the proposed codes over the conventional staircase codes can be much larger than the corresponding threshold gain in Table \ref{table2} because larger $(t_1,t_2)$ lead to lower miscorrection probability of iBDD.

\section{Stall Pattern Analysis}\label{sec:error_floor_analysis}
The error floor performance of the class of staircase codes is affected by stall patterns \cite[Def. 1]{Smith12}. To determine the BER due to stall patterns, we consider a fixed code block $\boldsymbol{B}_i$ and the error bits of stall patterns including positions in $\boldsymbol{B}_i$ and possibly additional positions in $\boldsymbol{B}_{i+1},\ldots$ but not in $\boldsymbol{B}_{i-1}$. The BER of the error floor is dominated by the occurrence probability of the stall patterns with the smallest size \cite{Smith12,7537380,8856224}. Consider a BSC with crossover probability $p$ and under miscorrection-free iBDD. The BER can be approximated by using the union bound technique $\mathsf{BER}_{\mathsf{floor}}\approx \frac{s_{\min}A_{\min}p^{s_{\min}}}{\text{size of} \boldsymbol{B}_i}$ following \cite[Sec. V]{Smith12}, \cite[Eq. (7.91)]{optical_FEC_2020}, where $A_{\min}$ is the multiplicity of minimum stall patterns, and $s_{\min}$ is the number of error bits of a minimum stall pattern.

In what follows, we present the main results for $s_{\min}$ to gain some insights into the error floor and justify the choice of code parameters. Due to space limitations, we omit the proof and the analysis of $A_{\min}$. For more details, we refer the interested reader to \cite{Min2022staircase}.

\begin{theorem}\label{prop:w2_smin}
Consider a SR-staircase codes with parameters $(t_1,t_2)$, $(q_1,q_2)$, and $w=2$. The exact number of the error bits of the minimum stall pattern is
\begin{align}\label{eq:smin_w2}
&s_{\min} = \min\bigg\{\max\left\{\left\lceil\frac{t_2+1}{q_1} \right\rceil  (t_1+1),\left\lceil\frac{t_1+1}{q_1} \right\rceil  (t_2+1)\right\}, \nonumber \\
&\max\left\{ \left\lceil\frac{t_1+1}{q_2} \right\rceil  (t_2+1),\left\lceil\frac{t_2+1}{q_2} \right\rceil  (t_1+1) \right\}\bigg\}.
\end{align}
\end{theorem}

Based on Theorem \ref{prop:w2_smin}, it is desirable to have $\max\{q_1,q_2\} \leq \min\{t_1,t_2\}$ when $w=2$ to ensure that any minimum stall pattern will not become a one-dimensional vector whose $s_{\min}$ becomes very small. Although the size of a minimum stall pattern for SR-staircase codes is smaller than that for the conventional staircase codes when both codes are with the same $(t_1,t_2)$, the proposed codes can still achieve a better error floor due to much smaller multiplicity [arxiv to be uploaded] and the use of BCH component codes with larger $(t_1,t_2)$.

For a large coupling width, we need to set $m_1 = m_2 \triangleq m$ and $q_1 = q_2 \triangleq q$ according to Sec. \ref{subsec:enc_w3}. In the interest of space, we consider $w \geq q+1$ in the subsequent analysis. Unlike for $w=2$, obtaining the exact $s_{\min}$ for $w>2$ is difficult. Instead, we find the lower bound on $s_{\min}$, which serves as the upper bound on the BER of the error floor.

\begin{theorem}\label{prop:largew}
Consider a SR-staircase code with parameters $(t_1,t_2)$, $m_1 = m_2 \triangleq m$, $q_1 = q_2 \triangleq q$, and $w \geq q+1$. The size of the minimum stall pattern satisfies
\begin{align}\label{eq:smin_w_large}
s_{\min}\geq  \frac{(\min\{t_1,t_2\}+1)(\min\{t_1,t_2\}+2)}{2}.
\end{align}
\end{theorem}

Notice that all of our designs in Table \ref{table2} satisfy $|t_1-t_2| \in \{0,1\}$ because these designs achieve a better threshold than those with $|t_1-t_2| >1$. Under this condition, the lower bound of $s_{\min}$ for $w\geq q+1$ in Theorem \ref{prop:largew} is larger than the exact $s_{\min}$ for $w=2$, $q_1 \geq 2$ and $q_2 \geq 2$ in Theorem \ref{prop:w2_smin}. Hence, the error floor can be improved by increasing $w$.

\begin{lemma}\label{lem:stall_w3}
Consider the SR-staircase code in Theorem \eqref{prop:largew} with $w\geq q+1$. If $(q,w,t_1,t_2)$ further satisfy one of the following conditions: 1) $ \min \{t_1,t_2\} \geq q$; 2) $\min\{t_1,t_2\}+1 \leq q $ and $ w \leq (\mathbbm{1}\{t_1\neq t_2\}+1)(\min \{t_1,t_2\}+1)$, then $s_{\min}$ is strictly larger than the lower bound in \eqref{eq:smin_w_large}.
\end{lemma}

Lemma \ref{lem:stall_w3} shows that if both $q$ and $w$ are not too large, the size of the minimum stall pattern can become larger. Hence, a proper choice of $(q,w,t_1,t_2)$ would lead to a better trade-off between threshold and error floor for SR-staircase codes.

\section{Numerical Results}\label{sec:sim}

We evaluate the performance of SR-staircase codes over the  additive white Gaussian noise (AWGN) channel. Note that all BCH component codes used in our designs do not have any extended parity bits.

\begin{figure}[t!]
	\centering
\includegraphics[width=3.1in,clip,keepaspectratio]{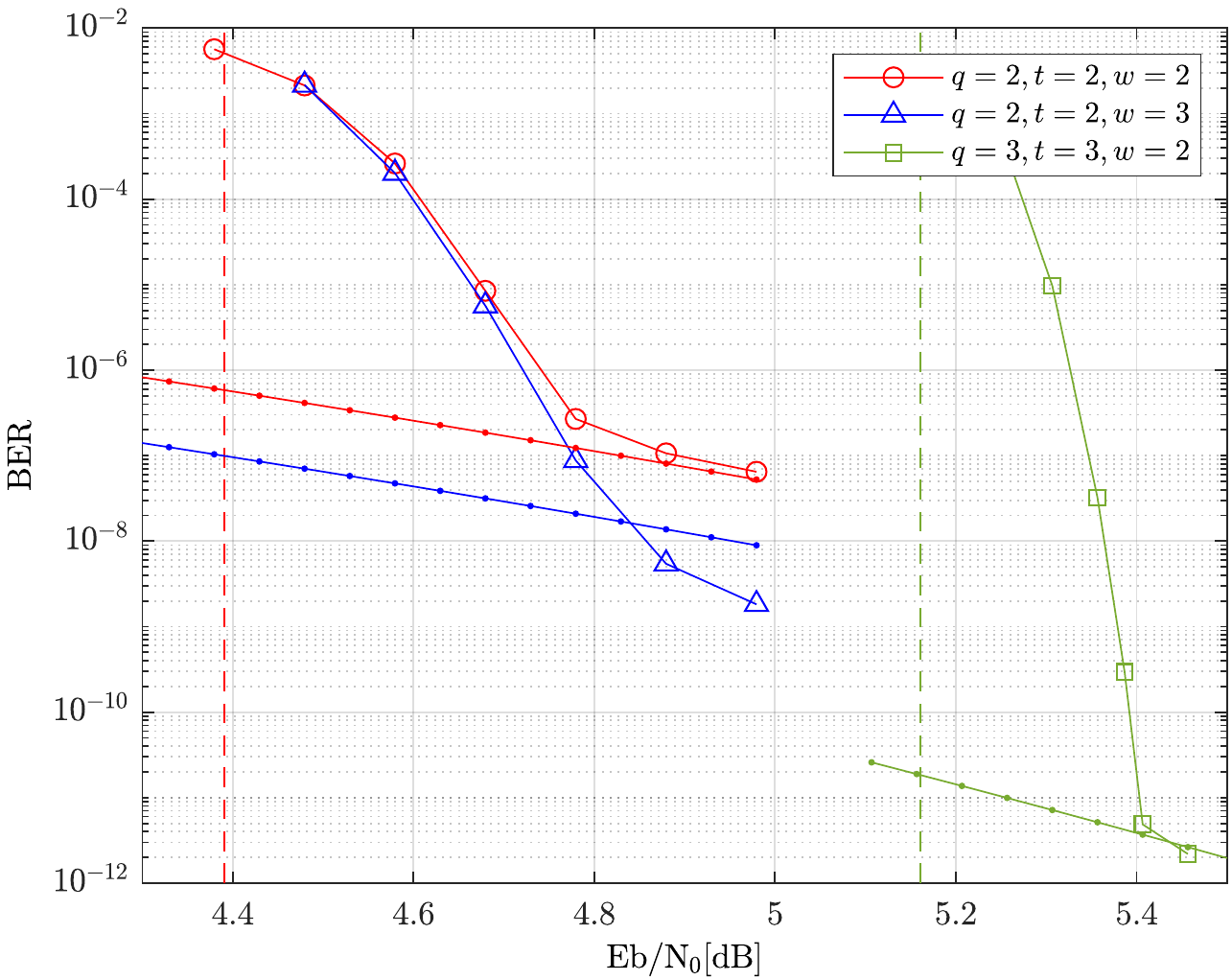}
\caption{Simulation results of SR-staircase codes to verify their decoding thresholds (dash line) and error floor (with marker `$\cdot$').}
\label{fig:BER_floor_v1}
\end{figure}

We first use simulation results to validate our theoretical analysis by assuming miscorrection-free iBDD. We construct three SR-staircase codes with parameters $(m,\nu,t,q,w)=(126,8,2,2,2)$, $(126,8,2,2,3)$, and $(441,9,3,3,2)$, respectively. The decoding window size is $W=7$. The simulated BER, decoding threshold (converted from the BSC threshold) and the estimated error floor $\mathsf{BER}_{\mathsf{floor}}$ are shown in Fig. \ref{fig:BER_floor_v1}. For the proposed codes with $w=2$, their simulated error floor BER matches closely to $\mathsf{BER}_{\mathsf{floor}}$ based on Theorem \ref{prop:w2_smin}. However, the $\mathsf{BER}_{\mathsf{floor}}$ for the code with $w=3$ can only be estimated based on Theorem \ref{prop:largew}. Since the code parameters $(t,q,w)=(2,2,3)$ satisfies the conditions in Lemma \ref{lem:stall_w3}, the actual error floor of the code with $w=3$ is lower than the $\mathsf{BER}_{\mathsf{floor}}$. Observe that the simulated waterfall performance for all the codes is also in agreement with the derived decoding threshold (the threshold curves for the codes with $t=2$ and $w\in \{2,3\}$ are overlapped). Therefore, both DE and error floor analysis can be used to effectively predict the simulated performance if the probability of miscorrection is low, which is the case in our subsequent design with large $(t_1,t_2)$.

\begin{figure}[t!]
	\centering
\includegraphics[width=3.1in,clip,keepaspectratio]{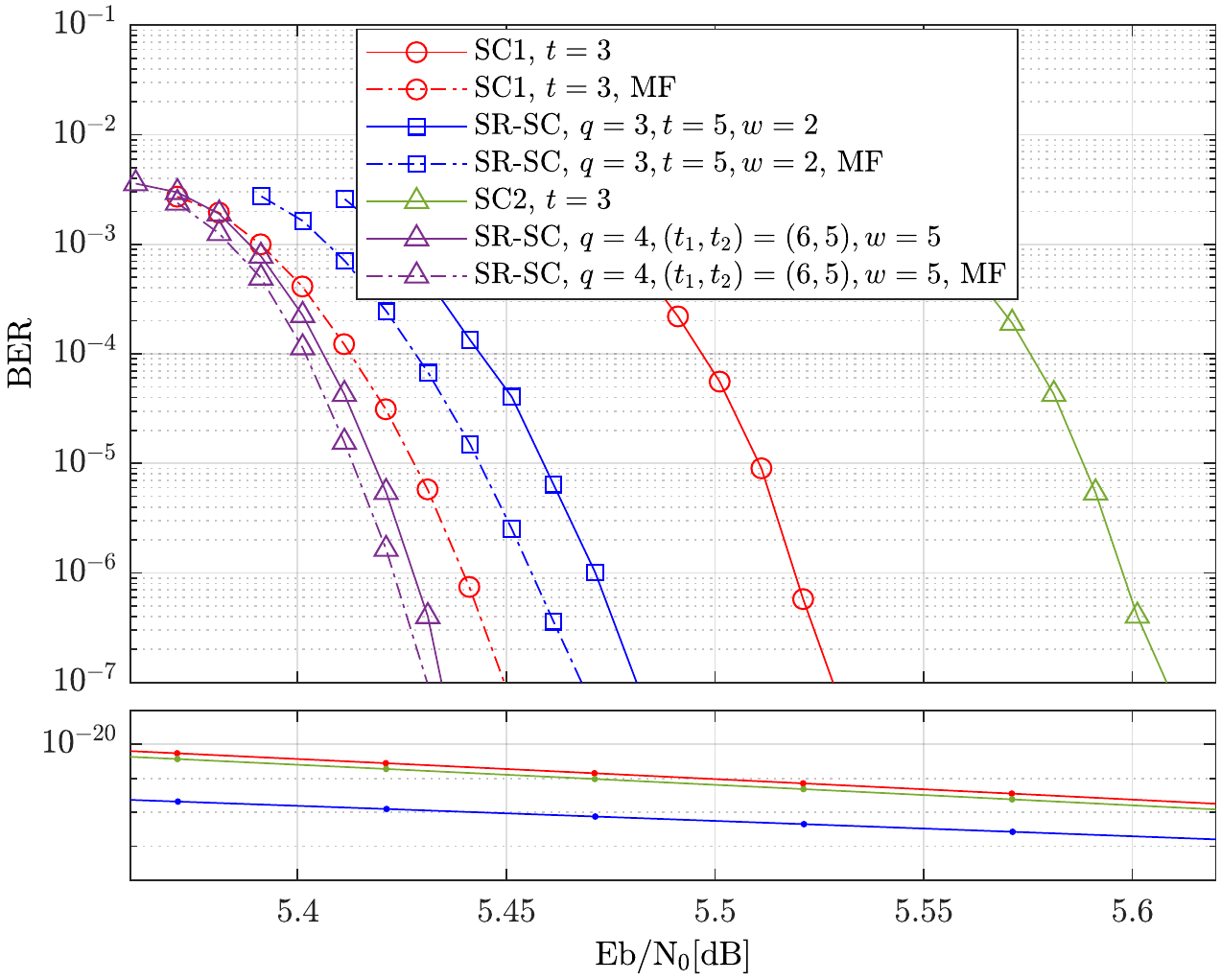}
\caption{BER and the estimated error floor of SR-staircase codes and the staircase codes in \cite[Sec. IV-C]{Smith12}.}
\label{fig:BER_v1}
\end{figure}

Next, we compare the designed SR-staircase codes with the conventional staircase codes. For SR-staircase codes (labeled as ``SR-SC''), we consider two designs from Table \ref{table2}, whose parameters are $(m,\nu,t,q,w)=(876,11,5,3,2)$, and $(m,\nu,t_1,t_2,q,w)=(964,11,6,5,4,5)$, respectively. We also consider two benchmark conventional staircase codes, where the first one (labeled as ``SC1'') has parameters $(m,\nu,t) = (510,10,3)$ and two parity bits extended following \cite[Sec. IV-C]{Smith12} while the second one (labeled as ``SC2'') has parameters $(m,\nu,t) = (478,10,3)$ and no extended parity bits. All the codes have rate 0.937 and comparable code block size as shown in Table \ref{table2}. The decoding window size is $W=9$. The error performance under iBDD (solid lines), miscorrection-free iBDD (dash lines, labeled as ``MF''), and the estimated error floor $\mathsf{BER}_{\mathsf{floor}}$ are shown in Fig. \ref{fig:BER_v1} (the $\mathsf{BER}_{\mathsf{floor}}$ of the SR-staircase code with $w=5$ is not shown in the figure as it is in the order of $10^{-33}$). Observe that SC2 under iBDD has the worst performance due to the highest probability of miscorrection. Even though SC1 uses two additional parity bits to reduce miscorrection probability, it still has a noticeable gap to its miscorrection-free performance. In contrast, all the proposed codes operate close to their miscorrection-free performance with iBDD and outperform the conventional staircase codes in terms of better waterfall and error floor performance. Most notably, the SR-staircase code with $w=5$ has the best performance among all the codes and achieves slightly better waterfall performance with iBDD than the convectional staircase code with miscorrection-free iBDD.

\section{Concluding Remarks}\label{sec:conclude}
We proposed SR-staircase codes, which are motivated and derived from the conventional staircase codes. The most appealing feature of the proposed codes is that one can employ stronger BCH component codes to construct a SR-staircase code to achieve better decoding threshold and error floor while having the same rate and similar block size compared to the conventional staircase code. The superior performance of the proposed codes over the conventional staircase codes was demonstrated via theoretical analysis and simulation.

\appendices

\bibliographystyle{IEEEtran}
\bibliography{MinQiu}

\end{document}